\documentclass[twocolumn,prb,aps,showpacs,floatfix]{revtex4}

\usepackage[dvips]{graphicx}

\usepackage{bm} 

\begin{document}

\title{Structural precursor to the metal-insulator transition in V$_2$O$_3$}

\author{P. Pfalzer}
\author{G. Obermeier}
\author{M. Klemm}
\author{S. Horn}
\affiliation{Institut f\"{u}r Physik, Universtit\"{a}t Augsburg,
Universit\"{a}tsstra{\ss}e 1, 86159 Augsburg, Germany }
\author{M. L. denBoer}
\affiliation{Queens College of CUNY, 65-30 Kissena Blvd., Flushing, New York 11367}
\date{\today}

\begin{abstract}
The temperature dependence of the local structure of V$_2$O$_3$ in the vicinity of the metal to insulator transition (MIT) has been investigated using hard X-ray absorption spectroscopy.
It is shown that the vanadium pair distance along the hexagonal c-axis changes abruptly at the MIT as expected. 
However, a continuous increase of the tilt of these pairs sets in already at higher temperatures and reaches its maximum value at the onset of the electronic and magnetic transition.
These findings
confirm recent theoretical results which claim that electron-lattice coupling is important for the MIT in V$_2$O$_3$.
Our results
suggest that interactions in the basal plane play a decisive role for the MIT and orbital degrees of freedom drive the MIT via changes in hybridization.
\end{abstract}


\maketitle

\section*{Introduction}

The metal to insulator transition in V$_2$O$_3$ is a long standing yet not completely resolved issue.
Originally being discussed as the classical example of a Mott-Hubbard transition,\cite{McWhan_69} this interpretation has to be questioned due to the intimate interplay of structural, electronic and magnetic changes in the phase diagram of this compound.\cite{McWhan_73}
Therefore it is not clear whether electronic correlations can completely account for the MIT in V$_2$O$_3$ or if structural effects play an important role.
The relevance of possible electron-lattice interactions has been stressed by recent theoretical work.\cite{Laad_05,Tanaka_02}

In an earlier paper\cite{Pfalzer_02} we have shown that a detailed understanding of the interplay of structural and electronic degrees of freedom requires both information on the global structure as measured by X-ray diffraction (XRD) and of the local structure of the compound.
Interatomic distances and symmetries can deviate locally from the global lattice symmetry, therefore indicating the presence of an order-disorder component of the transitions or of structural fluctuations.
Our earlier results demonstrate that the two insulating phases of V$_2$O$_3$, the paramagnetic insulating (PI) and the antiferromagnetic insulating (AFI) phase, are structurally very similar on a local scale and therefore indicate that there might exist a common route for both the transitions from the pa\-ra\-mag\-ne\-tic metallic (PM) phase to the PI and the AFI phase, respectively.

In this letter, we investigate the evolution of the local structure of V$_2$O$_3$ in the vicinity of the PM to AFI transition to shed light on the detailed structural changes associated with the MIT.
The results of our temperature dependent extended X-ray absorption fine structure (EXAFS) measurements show a rather complex behavior of the local structure, which differs along different crystallographic axes.
In particular, we observe a structural precursor to the MIT which consists of characteristic changes of nearest neighbor distances for the first time.

\section*{Experimental}
\medskip

EXAFS measurements were performed on a V$_2$O$_3$ single crystal grown by chemical transport using TeCl$_4$ as transport agent.
The crystal was mounted on a thin glass plate and reduced in thickness by lapping and polishing the sample down to approximately 12 $\mu$m which corresponds to roughly two absorption lengths.
This treatment allowed to detect the X-ray absorption in a transmission geometry, thereby avoiding damping of the signal due to selfabsorption effects which would have been present in fluorescence mode.\cite{Pfalzer_99}
The sample was oriented so that the angle between the polarization vector $\bm{E}$ of the incoming X-rays and the hexagonal $c$ axis
of the crystal
($\bm{c}_{hex}$) could be changed by rotating the sample around the surface normal.
Measurements were made with $\bm{E}$ parallel and perpendicular to $\bm{c}_{hex}$.
This procedure allows to separate scattering paths along and perpendicular to the hexagonal $c$ axis as detailed in ref. \onlinecite{Pfalzer_02}

The electrical resistivity was measured using a standard four-probe DC technique.
For each measurement the current was reversed to eliminate offsets.
Contacts were made by attaching platinum wires (diameter: 40~$\mu$m) to the sample by silver epoxy.
%
Metallic behavior was observed in the high temperature region, and a sharp jump in the resistance of more than three orders of magnitude at a temperature of 140 K signaled the MIT.\cite{cooling-note}
The slightly reduced transition temperature indicates that a small vanadium deficiency occurred during the growth of the single crystal.
The magnetic transition from PM to AFI occurred in a temperature region from 140 K down to 130 K as determined by magnetic susceptibility measurements using a Quantum Design superconducting quantum interference device (SQUID) magnetometer.

\begin{figure}
  \centering
  \includegraphics[width=8.3cm]{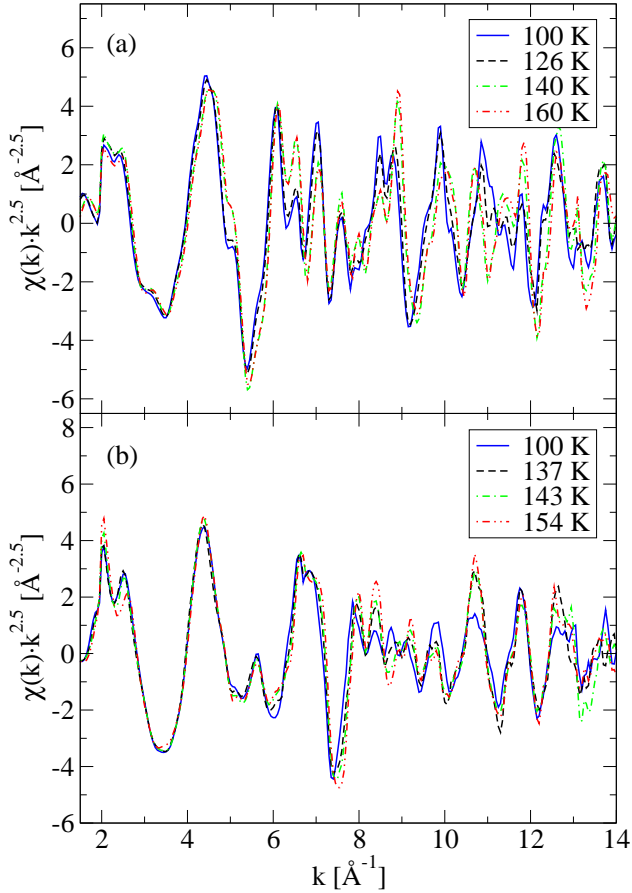}
  \caption{\label{ksp} (color online). EXAFS spectra of the V K edge of V$_2$O$_3$ in the vicinity of the MIT at T=140~K. The spectra in a geometry with $\bm{E}$ parallel to $\bm{c}_{hex}$ are almost identical within one phase and display an abrupt change at the MIT (a). Only smooth changes are observed in a setup where $\bm{E}$ is perpendicular to $\bm{c}_{hex}$ (b).}
\end{figure}

EXAFS measurements at the vanadium K edge (5465 eV) were performed at beam line X23B of the National Synchrotron Light Source at Brookhaven National Laboratory. 
Spectra at different temperatures above, at, and below the transition temperature were taken using a closed cycle helium cryostat. 
The sample was held under vacuum to avoid water condensation and to reduce heat transmission.
As no temperature sensor was available in the direct neighborhood of the sample, the resistivity of the crystal was monitored \emph{in situ} during the EXAFS experiment in the geometry with $\bm{E}$ perpendicular to $\bm{c}_{hex}$.
This allowed for a direct observation of the metal to insulator transition itself.
The electronic transition was observed at a temperature of 130 K at the cold finger of the cryostat.
The temperature difference of -10 K compared to the \emph{ex situ} resistance measurements can be attributed to a temperature gradient between the cold finger and the sample.
Temperatures given in the paper are corrected accordingly.

The crystal has been examined by XRD measurements after the EXAFS experiments were finished.
The investigations, both in $\Theta\,/\,2\Theta$ and in Laue geometries, confirmed that the MIT did not destroy the single crystal, as the diffraction patterns were those expected for a good single crystal and no signs of twinning could be observed.
%
In addition, spectra above the transition were measured before and after the crystal passed through the MIT during the EXAFS experiment.
No differences were observed in these spectra, ruling out the influence of possible cracks on the measurements as well.
%
Typcial spectra are shown in fig. \ref{ksp} and display that a good signal to noise ratio could be achieved at least up to a wavevector of $k=14$~\AA$^{-1}$ ($k$ being the wavevector of the outgoing photoelectron).

Standard EXAFS analysis using the ATHENA\cite{Ravel_05} and IFEFFIT\cite{Newville_01} programs was performed to extract path lengths and other parameters from the spectra.
The experimental spectra were normalized to the edge step and a smooth, atomic-like background was subtracted by minimizing the low-$R$ contribution in the Fourier-transformed data.\cite{Newville_93}
The EXAFS equation in cumulant expansion as defined in Ref. \onlinecite{Newville_95} and implemented in the IFEFFIT program was used up to the third cumulant $C_3$ to fit the experimental data to
model spectra calculated with FEFF8.2.\cite{Ankudinov_02}
For the Fourier transforms a window in $k$ space from 2 \AA$^{-1}$\ to 13 \AA$^{-1}$\ was used 
%
and the spectra were weighted by $k^{2.5}$.
Fitting was carried out
in real space in a range from 1.2 \AA\ to 4.1 \AA\ ($\bm{E} \parallel \bm{c}_{hex}$) and from 1.2 \AA\ to 3.65 \AA\ ($\bm{E} \perp \bm{c}_{hex}$), respectively.
The cluster of ions used for the calculation of the model spectra was derived from the lattice parameters and atomic positions resulting from XRD measurements for the trigonal structure of V$_2$O$_3$ at room temperature published in Ref. \onlinecite{Vincent_80}.

\begin{figure}
  \centering
  \includegraphics[width=8.3cm]{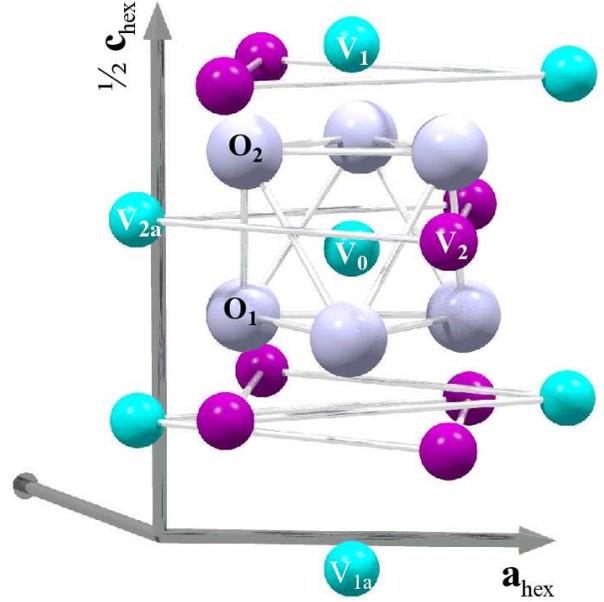}
  \caption{\label{structure} (color online). Room temperature structure of V$_2$O$_3$. Small (large) spheres represent vanadium (oxygen) ions. Only those oxygen ions which form an octahedron around the central vanadium ion V$_0$ are shown. Vanadium ions which have opposite spin in the AFI phase are drawn in different colors.}
\end{figure}

The structure refinement was performed by varying the interatomic distances to all vanadium ions up to a shell radius of 4.3 \AA, which corresponds to the V ions shown in fig.~\ref{structure}. 
In addition, backscattering from the six oxygen ions that form the coordination octahedron of the absorber was accounted for.
Inclusion of the second backscattering shell of oxygen ions improved the quality of the fit, but had no influence on the fitting parameters concerning the ions shown in Fig. \ref{structure}.
At the transition from the paramagnetic metal to the antiferromagnetic insulator, the lattice symmetry reduces from trigonal to monoclinic.
Comparing the interatomic distances of the trigonal and the monoclinic structure as calculated from the data published in Refs. \onlinecite{Vincent_80} and \onlinecite{McWhan_70,Dernier_70b,Calvert_91} 
and summarized in the two rightmost columns of table \ref{v2o3-V-dist}, 
it turns out that the transition increases the distance of the neighboring vanadium ions along the hexagonal $c$ axis (V$_0$ and V$_1$) by 0.04 \AA.
In addition, the threefold symmetry in the plane perpendicular to $\bm{c}_{hex}$ (the basal plane) is broken and one of the three vanadium neighbors in this plane (V$_{2a}$) increases its distance to the absorber (V$_0$) by 0.11 \AA, while the distance to the two other V$_2$ ions remains nearly constant.

\begin{table}
\begin{ruledtabular}
\begin{tabular}{cr@{$\,\cdot\,$}lr@{$\,\cdot\,$}lrr}
 & \multicolumn{4}{c}{$\Delta R$} & $R_{trigonal}$ & $R_{monoklin}$ \\ 
 & \multicolumn{2}{c}{$\bm{E} \parallel \bm{c}_{hex}$} & \multicolumn{2}{c}{$\bm{E} \perp \bm{c}_{hex}$} & & \\ \hline \\[-3ex]
V$_0$-V$_1$ & \multicolumn{1}{r@{}}{}&$dv1$ & \multicolumn{2}{c}{---} & 2.70 & 2.74 \\[1ex]
V$_0$-V$_{2a}$ & \multicolumn{2}{c}{---} & \multicolumn{1}{r@{}}{}&$dv2$ & 2.88 & 2.99 \\
V$_0$-V$_{2b}$ & \multicolumn{2}{c}{---} & \multicolumn{1}{r@{}}{}&\multicolumn{1}{r}{$0$} & 2.88 & 2.88 \\
V$_0$-V$_{2c}$ & \multicolumn{2}{c}{---} & $-2/11$&$dv2$ & 2.88 & 2.86 \\[1ex]
V$_0$-V$_{3a}$ & $-3/4$&$dv1$ & $-3/11$&$dv2$ & 3.47 & 3.44 \\
V$_0$-V$_{3b}$ & $-1/4$&$dv1$ & $-1/11$&$dv2$ & 3.47 & 3.46 \\[1ex]
V$_0$-V$_{4a}$ & $-6/4$&$dv1$ & $-6/11$&$dv2$ & 3.69 & 3.63 \\
V$_0$-V$_{4b}$ & $4/4$&$dv1$ & $4/11$&$dv2$ & 3.69 & 3.73 \\
V$_0$-V$_{4c}$ & $5/4$&$dv1$ & $5/11$&$dv2$ & 3.69 & 3.74 \\[1ex]
V$_0$-V$_{1a}$ & $-7/4$&$dv1$ & \multicolumn{2}{c}{---} & 4.30 & 4.23 \\
\end{tabular}
\end{ruledtabular}
  \caption{\label{v2o3-V-dist}Definition of the next neighbor V-V distances in the fitting model. The free parameters are $dv1$ for the change of the V$_0$-V$_1$ distance and $dv2$ for the change of the V$_0$-V$_{2a}$ distance, respectively. The changes $\Delta R$ of the remaining V-V distances used in the fits with with $\bm{E} \parallel \bm{c}_{hex}$ and with $\bm{E} \perp \bm{c}_{hex}$ are set as multiples of these values to represent the values calculated from the trigonal and monoclinic structure as measured by XRD (right columns) when $dv1$ and $dv2$ take on the respective values of $dv1=0.04$~\AA\ and $dv2=0.11$~\AA.}
\end{table}

To allow for this monoclinic distortion in the fitting model, the distance between the absorber and these V neighbors (V$_1$ for the spectra with $\bm{E} \parallel \bm{c}_{hex}$ and V$_{2a}$ for the spectra with $\bm{E} \perp \bm{c}_{hex}$) was allowed to vary freely, while the distances to the remaining V 
%
ions were given as multiples of this value
as defined in table \ref{v2o3-V-dist}.
This allowed to describe the changes of all V-V distances with a single parameter for each orientation ($dv1$ or $dv2$).
In the same way the positions of the O ions were split up as multiples of $dv1$ and $dv2$, respectively.
%
An additional isotropic change of the V-O distances was allowed to account for a possible change in the oxygen positions due to the vanadium deficiency of the crystal.
A damping term (Debye-Waller factor) was assigned to each neighboring shell, and for the nearest vanadium neighbors the third cumulant was used as an additional free parameter.
Altogether, 12 free parameters were used to fit the 16.7 independent points of each spectra.

\section*{Results}
\medskip

\begin{figure}
  \centering
  \includegraphics[width=8.3cm]{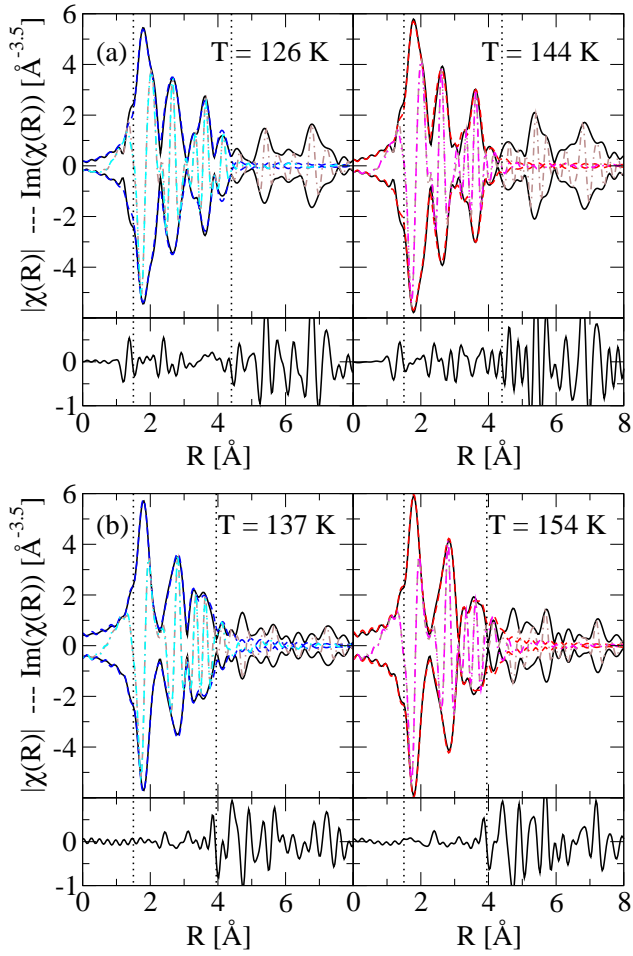}
  \caption{\label{rsp} (color online). Typical examples of the Fourier transformed EXAFS spectra of V$_2$O$_3$ at temperatures close to the MIT in a geometry with $\bm{E}$ parallel to $\bm{c}_{hex}$ (a) and with $\bm{E}$ perpendicular to $\bm{c}_{hex}$ (b). Solid and dash-dotted lines represent the amplitude and imaginary part of the Fourier transformed data, dashed and double-dotted dashed lines show the fitted model spectra. Data and fit coincide well in the fitting range (marked by vertical dotted lines) as can be seen from the difference spectra of the imaginary parts of data and fit in the respective lower panels. For this figure, the scattering phase shift of the central V ion has been corrected, which basically leads to a shift of the fourier transformed spectra and the fitting ranges to the right by $\approx 0.3$~\AA.}
\end{figure}

\begin{table}
\begin{ruledtabular}
\begin{tabular}{rrrrr}
Temperature [K] & \multicolumn{2}{c}{$\bm{E} \parallel \bm{c}_{hex}$} & \multicolumn{2}{c}{$\bm{E} \perp \bm{c}_{hex}$} \\
 & $r$-factor & $\chi_\nu^2$ & $r$-factor & $\chi_\nu^2$  \\ \hline \\[-3ex]
 40 & 0.020 & 197 & 0.004 & 56 \\
100 & 0.014 & 121 & 0.001 & 18 \\
122 & 0.014 &  25 & 0.006 & 17 \\
126 & 0.010 &  52 & 0.006 & 37 \\
130 & 0.009 &  47 & 0.002 &  7 \\
134 & 0.006 &  29 & 0.005 & 28 \\
137 & 0.007 &  24 & 0.005 & 46 \\
140 & 0.012 &  43 & 0.003 & 24 \\
143 & 0.013 &  58 & 0.004 & 33 \\
149 &       &     & 0.004 & 30 \\
154 &       &     & 0.003 & 34 \\
160 & 0.007 &  17 &       &    \\
165 &       &     & 0.003 & 21 \\
210 &       &     & 0.007 & 30 \\
\end{tabular}
\end{ruledtabular}
  \caption{\label{fit-quality}$r$-factors and reduced chi-squared ($\chi_\nu^2$) for the fits presented in figs. \ref{rsp}-\ref{sigma}.}
\end{table}

All measured spectra could be fitted successfully with a reliability factor of less than 0.02, indicating that the model was well suited to describe the data.
For temperatures slightly below and above the MIT, the Fourier-transforms of four spectra and the respective fits are shown in fig. \ref{rsp}.
The values for the reliability factors and reduced chi-squared for the fits are compiled in table \ref{fit-quality}.

\begin{figure}
  \centering
  \includegraphics[width=8.3cm]{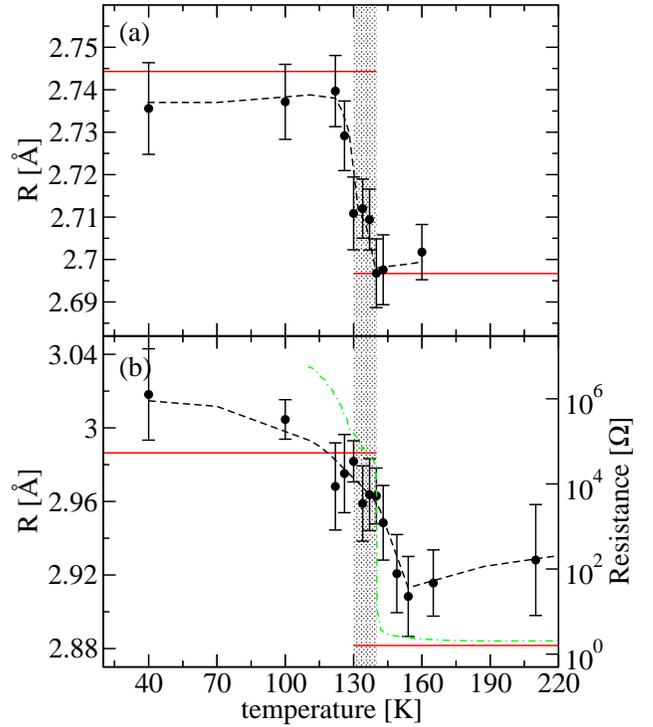}
  \caption{\label{dist} (color online). Temperature dependence of the distances of the V$_0$--V$_1$ (a) and V$_0$--V$_{2a}$ (b) scattering paths as resulting from the fits to the EXAFS spectra (dots). Solid lines represent the distances expected from XRD. The dash-dotted line in (b) shows the result of the \emph{ex situ} resistance measurement. The shaded temperature range marks the range of the magnetic transition. Dashed lines through the fit results provide a guide to the eye.}
\end{figure}

\begin{figure}
  \centering
  \includegraphics[width=8.3cm]{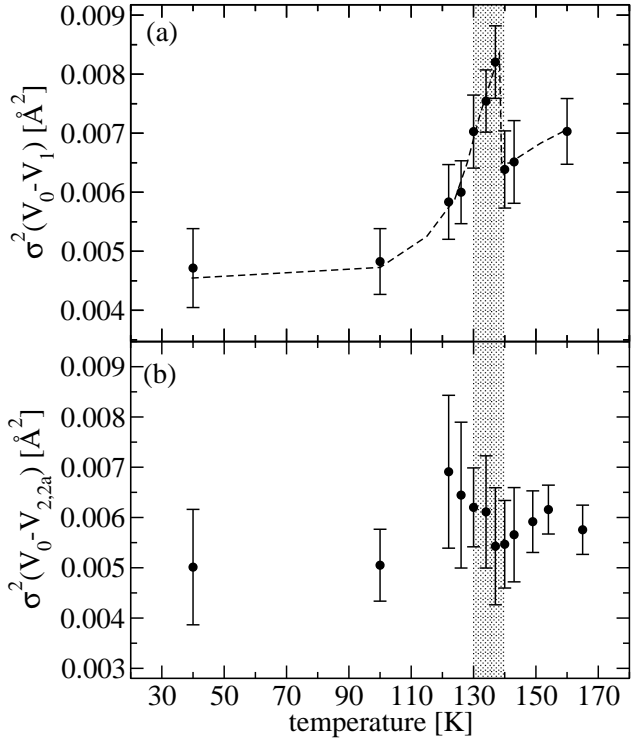}
  \caption{\label{sigma} Temperature dependence of the Debye-Waller factors $\sigma^2$ for the V$_0$--V$_1$ (a) and V$_0$--V$_{2,2a}$ (b) scattering paths as resulting from the fits to the EXAFS spectra. Like in Fig. \ref{dist}, the shaded temperature range marks the range of the magnetic transition. The dashed line through the fit results provides a guide to the eye.}
\end{figure}

Figures \ref{dist} and \ref{sigma} show the fit results for the scattering paths to the next vanadium neighbors along (V$_1$) and perpendicular (V$_2$, V$_{2a}$) to the hexagonal $c$ axis.
The geometry used in the measurement allowed for a separation of these paths: Backscattering from the V$_1$ neighbor only occurs in the spectra with $\bm{E} \parallel \bm{c}_{hex}$, while the V$_2$ and V$_{2a}$ backscatterers only contribute to the spectra with $\bm{E} \perp \bm{c}_{hex}$.
The Fourier transformed spectra show a well separated structure located between 2 \AA\ and 3 \AA\ 
(see fig. \ref{rsp}).
This structure is attributed to single scattering contributions either of the V$_1$ or the V$_2$, V$_{2a}$ scattering paths, only.
Therefore no correlations between fit variables of these paths can occur, although the distance from the absorber V$_0$ to all of these neighbors is similar.

Figure \ref{dist}(a) clearly shows that the distances between the next V neighbors along the hexagonal $c$ axis correspond to the values expected from XRD results:
The distance in the AFI phase at low temperatures is about 0.04 \AA\ larger than in the high temperature PM phase.
The change is abrupt as a function of temperature and occurs in a temperature window corresponding to that of the change in magnetic susceptibility.
The measurement uncertainty of the V$_0$--V$_1$ distances amounts to $\approx 0.01$ \AA.
The Debye-Waller factor $\sigma^2$, shown in Fig. \ref{sigma}(a), exhibits a sharp rise at the onset of the MIT, indicating either large fluctuations or a static distribution of widely varying distances while the magnetic transition occurs.

The next neighbor distances in the basal plane display a very different behavior.
First of all the distance from the absorber ion V$_0$ to the V$_{2a}$ neighbor is larger than expected from XRD in the high temperature PM phase [see Fig. \ref{dist}(b)], i.e.\ larger than the V$_0$--V$_2$ distances.
This break of the threefold symmetry expected from the long range structure probed by XRD corresponds to a monoclinic symmetry.
The elongation of the V$_0$--V$_{2a}$ distance
, which could be determined with an accuracy of $\approx 0.02$ \AA,
measures the size of this monoclinic distortion.
In the PM phase it is reduced to about 30 \% of the distortion in the AFI phase, as was first observed in Ref. \onlinecite{Frenkel_97}.
The transition to the full monoclinic distortion in the basal plane is much broader in temperature than that observed for the V-V distance along the hexagonal $c$ axis.
The onset occurs in a region well above the electronic and magnetic transitions.
This is consistent with the fact that the Debye-Waller factor of the in-plane V neighbors [Fig. \ref{sigma}(b)] does not show a significant change at the MIT.

Third cumulants 
($C_3$ or $\sigma^{(3)}$) 
which account for the anisotropy of the scattering paths to the next V neighbors where found to have a significant influence on the spectra in the low temperature AFI phase, but are negligible in the PM phase. 
The third cumulants show only a weak temperature dependence in the AFI phase and adopt values of $\approx -0.0003$ \AA$^3$ for the V$_0$--V$_1$ scattering path and $\approx -0.0007$ \AA$^3$ for the V$_0$--V$_2$ and V$_0$--V$_{2a}$ scattering paths, respectively.
The error is estimated to approximately $\pm$ 25 \%.

Distances between the central vanadium ion V$_0$ and the oxygen ions forming its coordination octahedron did not deviate from the values expected from XRD within error bars, i.e.\ these vanadium-oxygen distances remain nearly constant throughout the whole temperature range.

Summarizing, we can characterize the structural evolution at the metal to insulator transition of V$_2$O$_3$ as follows:
Structural changes along the hexagonal $c$ axis are restricted to a sharply bounded temperature range given by the MIT.
Perpendicular to $\bm{c}_{hex}$, however, the tilt of the $c$-axis V pairs, indicated by the elongation of the V$_0$-V$_{2a}$ bond, starts to increase already far above the MIT.

\section*{Discussion}
\medskip

Precursor effects to the MIT in V$_2$O$_3$ have already been observed in electrical resistance under pressure\cite{Carter_94} and very recently in the sound velocity measured for thin films.\cite{Mueller_05}
%
The results of our local structure determination show for the first time that structural changes preceed the changes in transport and magnetic properties.

The analysis of the polarization dependent measurements reveals that two distinct movements with different temperature dependencies have to be distinguished: 
The positioning of the vanadium ions within the basal plane and the change of the V-V pair distance perpendicular to the basal plane, i.e. along the hexagonal $c$-axis.
The former movement determines the rotation of these vanadium pairs, ranging from an alignment parallel to $\bm{c}_{hex}$ in the PM phase to a tilt of roughly 2$^\circ$ away from this axis in the AFI phase.\cite{Dernier_70b}

Following Tanaka,\cite{Tanaka_02} hybridization between the $a_{1g}$ and the $e_g^\pi$ orbitals increases on rotating the V$_0$--V$_1$ pairs out of the direction of the $c$-axis.
Depending on the hybridization strength, two minima of the free energy exist, which are characterized by two different equilibrium distances between the two nearest V neighbors along $\bm{c}_{hex}$.
While the electron-lattice coupling is weak when the V-V pair is aligned parallel to the hexagonal $c$-axis, increased hybridization on tilting the pair renders the orbital configuration almost energetically degenerate.
The so modified interactions favor a new ground state configuration with a reduced occupation of the $a_{1g}$ orbital.\cite{Park_00, Mueller_97b}
In this configuration, the electron-lattice coupling is strong and an enlarged V pair distance is expected.

We observe a continuously increasing tilt, starting well above the MIT and continuing even within the AFI phase. 
Along with it, the hybridization strength increases as just pointed out.
At a certain hybridization strength, the system snaps into the second minimum of the free energy, which corresponds to the enlarged pair distance.
Therefore it is plausible that the V$_0$--V$_1$ distance expands primarily after the tilt of the pair has adopted its maximum value,
what is exactly the behavior we observe experimentally.
Our measurements thereby substantiate the scenario proposed by Tanaka which does not regard the metal-insulator transitions in V$_2$O$_3$ (both the PM to AFI and the PM to PI transitions) as usual Mott transitions.\cite{Tanaka_02}
The fact, that on a local scale the trigonal symmetry is not fully recovered in the metallic phase, suggests that the trigonal and monoclinic structure are very close in energy, allowing fluctuations between the two.
We want to point out that the present results are consistent with our earlier findings\cite{Pfalzer_02} that the trigonal symmetry can be broken on a local scale while the global symmetry remains trigonal.

In step with the separation of the two neighboring V ions along the hexagonal $c$ axis an ``umbrella like'' distortion of the oxygen coordination octahedron around each V ion occurs.\cite{Dernier_70a}
Thus, the trigonal distortion of the coordination octahedron is directly determined by the distance between the next V neighbors along $\bm{c}_{hex}$.
The increased distortion of the oxygen octahedron is responsible for the stabilization of the insulating state\cite{Laad_05} since it results in a large shift of (dynamical) spectral weight.

However, from theoretical considerations it has been pointed out\cite{Elfimov_03} that in-plane interactions are important for the MIT in V$_2$O$_3$ and and should not be treated as small perturbations.
This view is supported by our measurements which suggest that in-plane interactions are a decisive ingredient to the MIT by triggering the modification of the V$_0$--V$_1$ distance via changes in hybridization.

In this context it is interesting to note that the Debye-Waller factor of the in-plane vanadium neighbors does not display substantial changes at the transition.
If structural degrees of freedom were the driving force behind the increase of the in-plane interatomic distances, the Debye-Waller factor should be strongly affected.
A possible explanation could be that the change in hybridization is driven by orbital interactions.
The change in hybridization will also result in a change of the crystal field splitting and accordingly alter the orbital occupation.
Evidence for orbital interactions is given from neutron scattering results by Bao \emph{et al.}\cite{Bao_97}
As shown by Laad \emph{et al.}, the variation of orbital occupation can result in a first order phase transition.\cite{Laad_03}

\section*{Conclusion}
\medskip

In conclusion, the measurements of the local structure of pure V$_2$O$_3$ in the vicinity of the metal to insulator transition show that complex changes of the interatomic distances on a local scale are decisive for the changes of the physical properties at the transition.
We can distinguish between two structural effects with different temperature characteristics:
At the onset of the electronic and magnetic transition temperature, the next vanadium neighbors along the hexagonal $c$ axis 
abruptly
increase their distance, resulting in an increase of the
trigonal crystal field component and a shift of spectral weight, rendering the system insulating.
However, a
continuous
increase of the monoclinic distortion in the basal plane is already observed far above the MIT reaching its maximum value at the onset of
the electronic and magnetic transition.
While the monoclinic distortion, in other words the tilt of the $c$-axis vanadium pair against this axis, determines the orbital configuration via changes in hybridization strength, the pair distance only reacts to these changes by locking into one of the two minima of the free energy defined by the orbital configuration.
These findings suggest that the metal to insulator transition could be driven by changes in the hybridization triggered by in-plane orbital interactions.

\begin{acknowledgments}
We appreciate valuable assistance in the measurements from J.\ Kirkland at NSLS.
This work was supported by the DFG through SFB 484.
The NSLS is supported by the DOE.
\end{acknowledgments}

%

\begin{thebibliography}{25}
\bibitem[{coo()}]{cooling-note}
\bibinfo{note}{{All temperatures given in this letter are measured on
  cooling.}}
\end{thebibliography}
%


\end{document}